\documentclass[sigconf]{acmart}
\graphicspath{{Figures/}}
\usepackage{todonotes}
\usepackage{xspace}
\usepackage{wrapfig}
\newcommand{\changed}[1]{#1}
\AtBeginDocument{%
  \providecommand\BibTeX{{%
    \normalfont B\kern-0.5em{\scshape i\kern-0.25em b}\kern-0.8em\TeX}}}

\copyrightyear{2023} 
\acmYear{2023} 
\setcopyright{acmlicensed}\acmConference[CHI '23]{Proceedings of the 2023 CHI Conference on Human Factors in Computing Systems}{April 23--28, 2023}{Hamburg, Germany}
\acmBooktitle{Proceedings of the 2023 CHI Conference on Human Factors in Computing Systems (CHI '23), April 23--28, 2023, Hamburg, Germany}
\acmPrice{15.00}
\acmDOI{10.1145/3544548.3581363}
\acmISBN{978-1-4503-9421-5/23/04}

\begin{document}

\title{Message Ritual: A Posthuman Account of Living with Lamp}

\author{Nina Rajcic}
\orcid{0000-0001-6501-5754}
\affiliation{%
  \institution{SensiLab, Monash University}
  \streetaddress{900 Dandenong Road}
  \city{Caulfield East}
  \state{Victoria}
  \country{Australia}
  \postcode{3145}
}
\email{Nina.Rajcic@monash.edu}

\author{Jon McCormack}
\orcid{0000-0001-6328-5064}
\affiliation{%
  \institution{SensiLab, Monash University}
  \streetaddress{900 Dandenong Road}
  \city{Caulfield East}
  \state{Victoria}
  \country{Australia}
  \postcode{3145}}
\email{Jon.McCormack@monash.edu}

\renewcommand{\shortauthors}{Rajcic and McCormack}
\renewcommand{\MR}{\emph{Message Ritual}\xspace}

\begin{abstract}
    As we become increasingly entangled with digital technologies, the boundary between human and machine is progressively blurring. Adopting a performative, posthumanist perspective resolves this ambiguity by proposing that such boundaries are not predetermined, rather they are enacted within a certain material configuration. Using this approach, dubbed `Entanglement HCI', this paper presents \emph{Message Ritual} -- a novel, integrated AI system that encourages the re-framing of memory through machine generated poetics. Embodied within a domestic table lamp, the system listens in on conversations occurring within the home, drawing out key topics and phrases of the day and reconstituting them through machine generated poetry, delivered to household members via SMS upon waking each morning. Participants across four households were asked to live with the lamp over a two week period. We present a diffractive analysis exploring how the lamp \emph{becomes with} participants and discuss the implications of this method for future HCI research.
\end{abstract}


\begin{teaserfigure}
  \includegraphics[width=\textwidth]{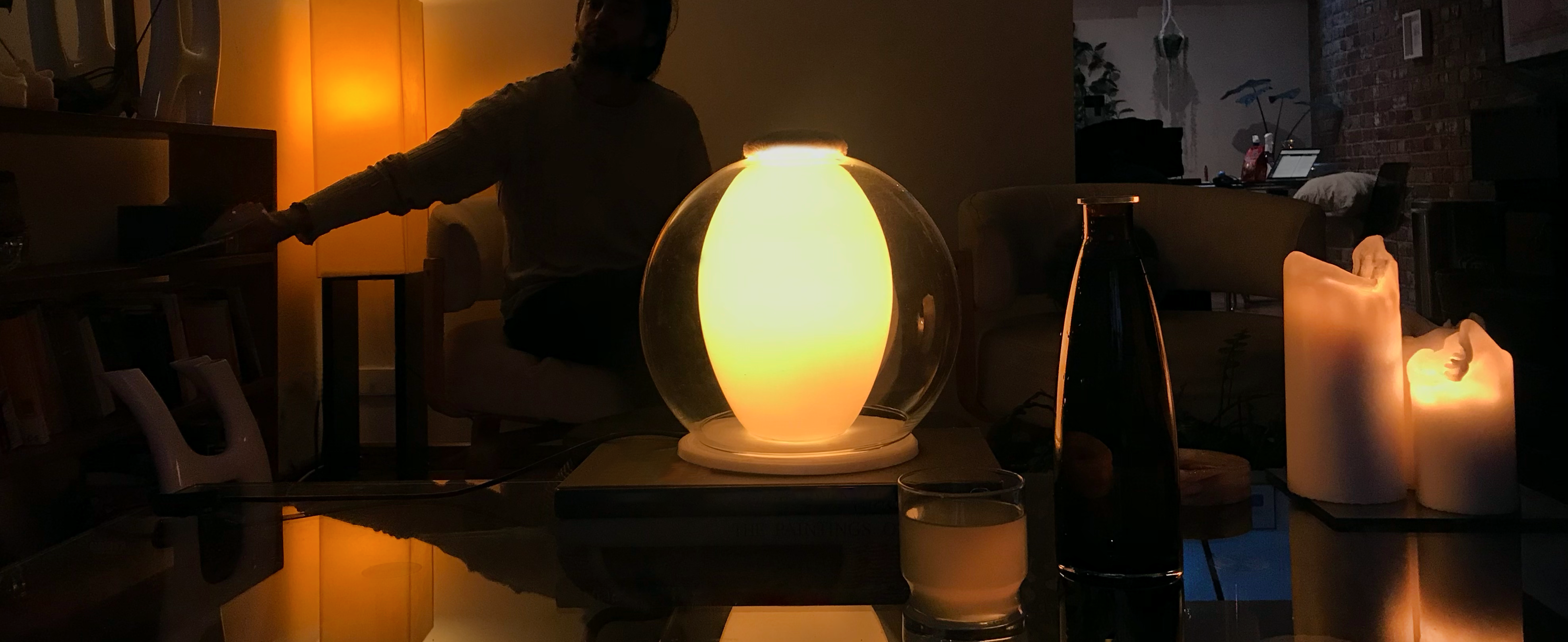}
  \caption{The \emph{Message Ritual} lamp.}
  \Description{The Message Ritual lamp in a domestic setting on a coffee table.}
  \label{fig:lamp}
\end{teaserfigure}

\maketitle

\section{Introduction}
\label{s:introduction}
\begin{quote}
    Rituals serve as a background against which our present times may be seen to stand out more clearly.\
\flushright --- Byung-Chul Han \cite[p. vi]{Han2020}
\end{quote}

As we move into a new age of Artificial Intelligence (AI) increasingly dominated by `foundation models' \cite{bommasani2021opportunities}, technology is broadening its role from that of being a tool or providing a service, to one of being an interlocutor, companion or co-creator. For many of us, our increasingly deep and nuanced entanglements with technology mark the beginning of a new era of human-machine \emph{relationships}\footnote{or perhaps more cynically, dependencies.} \cite{Bickmore2005, Benyon2011}.
This paper offers a practical, affirmative vision of one possibility for what our encounters with technology could become. We come from a design perspective, motivated by recent movements in HCI and posthumanism \citep{Wakkary2021,Forlano2017}. Recognising that technology is not benign, and that by incorporating emerging AI models it carries increasingly powerful agency \cite{Tholander_2012,Knappett:2008aa,Rosiek2020}, we shift from the rubric of human-centred design to that of human and non-human entanglements, with an emphasis on agential relations and the transformative nature of \emph{things} \cite[p95]{Wakkary2021}. Our research is motivated by an exploration of the possibilities for new, open-ended relationships we might \emph{want} to have with AI technology, rather than designing to solve a specific problem or engineer a particular outcome.

We present the design of a new kind of domestic technological artefact: \MR, an ambient AI system embodied in the form of a table lamp. The lamp listens in to household conversations, and from them derives important topics and themes. These themes are used as the basis for a unique machine-generated narrative that is delivered daily via SMS to each household member. \changed{\MR is an experimental and bespoke artwork that forms part of a broader thesis exploring the role of ritual in AI technologies.}
Four different households were recruited to live with the lamp for an extended period. We present a diffractive analysis on the system and its entanglements with both participants and researchers. \changed{This analysis revealed that amongst the participants there was no singular conclusion or way to understand the lamp, but rather multiple, sometimes contradictory insights arising from their entanglements with the system. We use this  approach to demonstrate how the adoption of posthumanist thinking is valuable when designing novel technological artefacts, and as a methodological framework for exploring a design's entanglements with both the human and non-human.}

\section{Background and Related Work}
\subsection{Human Machine Relationships}

In the seminal 1960 paper \emph{Man-Computer Symbiosis} \cite{licklider1960man},
Licklider envisions a world in which humans and computers are intimately coupled, forming a symbiotic relationship from which emerges a new kind of computer, and a new kind of human. Licklider's formative musings, among others, became the foundation of the field of Human-Computer Interaction (HCI) as we know it today. With the advent of the Personal Computer (PC), there was a shift from simple task-based processing towards real-time interfaces between human and machine.

Computers first became commonplace in the office environment, and so early HCI research concerned itself with achieving goals of functionality, productivity and efficiency. Consequently, the methodologies adopted during this time were those aligned with task-based, goal-oriented evaluation, largely stemming from classical cognitivism \cite{harrison2007three}.
As computing permeated from the workplace into the home, then from the home onto the body, HCI research began to explore more nuanced, subjective and distinctly human needs such as emotion, creativity, and loneliness \cite{bodker2006second, bodker2015third}. The epistemological processes within HCI research likewise shifted from engineering-inspired positivism towards socially constructed and situated forms of knowledge production.

Today, the intimate couplings between people and their personal devices are widely evident. We are, at all times, surrounded by a network of interconnected devices that monitor, mediate, and mould our parallel and hyperconnected realities. We have become entangled with technology in such a way that it's difficult to draw the boundary between human and machine. AI, brain-computer interfaces, and bio-implants are just a few of the technologies currently gaining prominence that challenge our existing conceptions of agency, autonomy, and the interface. Simultaneously, HCI research is on the cusp of it's fourth wave as the traditional notions of the `user' and the `interface' are breaking down \cite{baumer2017post}. Frauenberger proposes that in order to settle its ``epistemological troubles, the ontological uncertainties and the ethical conundrums,'' HCI research must now turn to posthumanist, or entanglement theories \cite{frauenberger2019entanglement}. 

Posthumanist philosophies, in the broadest sense, call for a radical de-centering of the human, marked by a return to materiality \cite{braidotti2019posthuman}. 
With early roots in Donna Haraway's conceptualisation of the cyborg \cite{haraway2006cyborg}, posthumanist thinking puts into question the foundations laid by the secular humanist movement%
---that humans embody the locus of agency, morality, rationality, and individuality. With the accelerating growth of technology and its increasing integration into daily life, the very definition of `human' becomes strained, the separation between `human' and `non-human' begins to break down, and a dualist understanding of the world (e.g. nature/culture, subject/object, mind/body) no longer serves the reality before us. Instead, relational ontologies are adopted as the new paradigm, proposing that the relations between things are more fundamental ontologically than the things themselves \cite{Schaab2013}.

\subsection{Posthumanism and HCI}

The central tenets of posthumanism, when applied to HCI research and design practices, put into question a number of the conceptual foundations currently in place. Firstly, the notion of the `user', who is often perceived as homogeneous, or at best falling within a handful of simplistic categories based around perceived skills, gender, age, etc. This conceptualisation not only reduces the human to a prescribed stereotype, but suggests that only the human possesses and exerts agency over the machine. Secondly, one must ask where exactly lies the `interface', when moving away from traditional screen-based technologies, and towards embedded and embodied computing. Adopting relational ontologies leads us to the understanding that in the designing of `interfaces', we are, in fact, designing the relationships between human and machine through which humanity continues to be shaped \cite{rosenberger2015postphenomenological}. 

Barad's account of performativity as a metaphysics \cite{barad2003posthumanist} offers further insight into how we might approach this notion of human-machine relationships. The boundaries between human and machine are not predetermined, instead they are \emph{enacted}. To draw these lines too early in the design process is to neglect the dynamic nature in which a human and machine will negotiate their respective roles, responsibilities, and agencies. Rather than designing interactions with fixed roles set up to be assumed by `user' and the machine, we can instead design for the relationship as a whole. This involves taking into account the user during their active engagement with an artefact along with the greater context of the situated relationship, and moreover how it develops over time. For example, the effect on participants when they're `not using' the artefact, as well as the effect on the relationships between human participants, participant and designer, and those of participants within their wider networks. These entanglements should be recognised during the entirety of the design and evaluation processes.

Agential cuts must necessarily be enacted at one scope or another---the methodology does not feign to account for the `whole', but rather to take accountability of these cuts by explicitly acknowledging when and where they are made, and ultimately to take a multiplicity of configurations into consideration when trying to better understand the effect of bringing a novel artefact into the world. To bring into contrast this multiplicity of states is what Barad would call \emph{diffraction as a methodology}. Once again utilising the metaphor of the photon, Barad proposes that diffraction leads to a ``mapping of the effects of differences'', as opposed to reflection in which ``much like the infinite play of images between two facing mirrors, the epistemological gets bounced back and forth'' \cite{barad2003posthumanist}. Diffraction as a methodology is likewise proposed by Frauenberger \cite{frauenberger2019entanglement} as a valuable tool in HCI research---rather than asking what makes an artefact `work', we should instead ask how an artefact becomes `different things'.

\subsection{Habit and Ritual}

Perhaps one of the most prominent contemporary relationships we observe today is that between a person and their smartphone. Although the ambitions of artificial general intelligence (AGI) research still sit firmly at the horizon, we have already formed relationships with AI systems that have a profound impact on everyday life. Social media platforms have adopted AI for numerous applications, but most notable are the AI systems that curate, in real-time, the stream of content that each user is exposed to, known to us simply as `the Feed'. The goal of these algorithms is to increase engagement, and ultimately to maximise screen-time. The result is an invisible yet highly targeted system that affects behaviour change---working most effectively when the user is unaware.

Yet the presence of agency within these algorithms is unmistakable. Social networks provide the ideal example of an assemblage of human and non-human agents, illustrating the posthuman reality of today. The algorithms are distinctly inhuman; they are indifferent to the humanist values of sovereignty, freedom, self-determination, and liberty. They are indifferent to whether or not the end-user is human at all. Social media feeds favour \emph{immediacy,} transfiguring time in such a way that the past is to be forgotten, and the future cyclic and predictable. Similarly, spacial locality is reconfigured, with information `depart[ing] from anywhere, and arriv[ing] everywhere' \cite{dejong2019posthuman}.

The disruptive nature of these non-human agents on human perceptions and behaviours can be observed across many aspects of society. %
In the case of social media, the favoured behaviour of the algorithms, and hence the generated behaviour in users, is one of unreflective, reactionary engagement. Users are rewarded for their unceasing engagement through clicks, likes, comments, shares---in addition to the infinite scroll of hyperpersonalised content. The relationships engendered by social media networks are then, \emph{by design}, habitual and largely unconscious. Habits are, in the broadest definition, automatic and repetitive behaviours that are learned and enforced through action \cite{neal2006habits}. Habits can be both positive and negative, yet in any case markedly require little conscious thought or intention, often becoming `second nature'. Rituals similarly involve routines, but can be differentiated by their intentionality, deep emotional involvement, and their `socially shared meaning' \cite{giovagnoli2020habits}. Notably, rituals serve a societal function, in that they communicate and reinforce the shared values and identity within a culture \cite{rao2006ritual}.

The idea of incorporating rituals into technological design is not new (see, e.g.~\cite{Domingues_1999, Gaver2010, Sas_2016, Kirk2016, Chatting2017, Mah_2020, Kluber2020}). In some cases the aim has been to re-establish traditional or existing rituals in a new technological form, in others new rituals are developed in response to problems proselytised by technology. Kl\"{u}ber and colleagues \citep{Kluber2020} designed a tangible artifact for relationship transition rituals, demonstrating that tangible devices can shape new ritual experiences and provide new ways of dealing with relationship uncertainty. Browne and Swift \citep{Browne2018} employ techniques of defamiliarisation and performance art to recontextualise and critique neural networks and AI more generally via a form of Ouija board. Work by Chatting and colleagues \citep{Chatting2017} developed a bespoke set of devices that permitted parents separated from their son to create a ritual of location sharing and reflective communication. A ``telescope'' device and accompanying ``totem'' allowed the child's parents to communicate their current remote location, which the child could find via the telescope. Sas and colleagues \citep{Sas_2016} explored the rituals of ``letting go'' (of people or material objects) used by psychologists and applied them to the letting go of digital artefacts, through a framework that highlighted temporality, visibility, and force.  

While modern human societies often seek to re-establish past rituals that have been displaced or subsumed by technological development, the philosopher Byung-Chul Han offers an alternate view on the role and value of rituals in contemporary technological cultures. He sees them as symbolic acts, constituted through symbolic perception, that bring forth `community without communication' \cite{Han2020}. Seeing the world as data rich, but symbol poor, Han argues that the value of rituals is in their ability to be `symbolic techniques of making oneself at home in the world'. That is, rituals are to time what the home is to space: they render time \emph{habitable} \cite[p.7]{Han2020}. In a world obsessed with relentless consumption, we are surrounded by disappearance---a destabilising force on life. Rituals offer an alternative to this, by \emph{lingering} they can endure, making time accessible, structuring time and furnishing it, like a home.

\subsection{Related HCI Design Work}

Several elements of the work presented here have been previously explored in design and HCI research in different contexts. Strong and Gaver first proposed the idea of simple domestic objects with embedded technology to evoke non-verbal remote communication and intimacy \cite{Strong1996}. Nowacka and colleagues \cite{Nowacka2018} developed an interactive lamp for the home that responded to movement gestures as a means of communication. Kirk and colleagues \cite{Kirk2016} focus of the use of phatic communication, using what they term ``phatic devices'' to support new forms of communication for mobile and remote workers using a research through design approach \cite{Frayling93,Gaver2012}. These projects use technology to replace a lost intimacy or communication brought about by physical separation. In contrast, our work seeks to reframe individual memory through AI generated poetic interpretations of personal conversations. Rather than replacing something lost through technology, \MR asks something of those that it is embedded with via the ritual of human communication, and in return it offers machine interpreted reflections back to those who have participated in this ritual.

\changed{Other work in HCI research has experimented with ritualising our interactions with digital assistants. Benabdallah's \textit{Sybil} presented a glowing orb with attached breathing sensor, the orb asking participants to synchronise their breathing with it's pulsating light \cite{Benabdallah2020}. The system then speaks a short, machine-generated ``prohecy'' based on the quality of the breathing patterns. The ``Voices and Voids'' project subverted a traditional voice assistant system through art experiments to reveal their hidden characteristics and inner workings \cite{Desjardins2021}. The researchers looked at transcripts from voice assistants and derived a series of performative vignettes based on issues of misinterpretation, mistakes, raising issues of ethics and privacy. In a design-led inquiry, Parviainen and Søndergaard experimented with whispering as a way of interacting with personal voice assistants \cite{Parviainen2020} examining how this mode of interaction might facilitate the experiential qualities of creepiness, trust, and intimacy. Our project uses a similar design-led enquiry, however in contrast to these projects \MR is not positioned as a voice assistant -- it doesn't speak, nor does it acknowledge or respond directly to spoken conversation. Additionally, any interactivity with the system is deliberately deferred: unlike these previous projects there is no immediate response or reward, one must wait until the next day before the receiving a text message from the lamp.}
 
Related to the design approach used here are ``mediating'' approaches, such as \emph{postphenomenology}, which views technology as a transformative mediator of human experiences and practices rather than an inert functional object \cite{Ihde1993,Verbeek2005,Verbeek2015,Hauser2018}. Verbeek \citep{Verbeek2015} suggests that designers ``do not merely design products, but human practices and experiences'' and therefore, ``Designing things is designing human existence''. Thus, humans and technology co-constitute human subjectivity and our objective experience of the world \cite{Rosenberger2015,Hauser2018}. Projects such as ``Tilting bowl'' (a bowl that randomly tilts, given to a small group of philosophers) \cite{Wakkary2018}, ``Morse Things'' (networked ceramic bowls that communicate via Morse code, deployed to the households of six interaction designers)  \cite{Wakkary_2017} typify this postphenomenological approach to HCI design. 

In a recent book, Ron Wakkary outlines the role of posthumanist thinking for design, including design in HCI \cite{Wakkary2021}. He argues that the main arc of design over the last forty years has been to prioritise human values, conceptualised anthropocentricly through a series of paradigms such as human factors, ergonomics, embodied interaction and human-centred design. While such paradigms have served us well, they have arguably come at a huge environmental, social and cultural cost. Posthumanist thinking offers a potential way out of this crisis, requiring greater humility in design, shifting the focus away from `the power of self-reflexive human reasoning to situated, partial, and multiple ways of knowing.' \cite[p.2]{Wakkary2021}. Posthuman design explores what it is like to \emph{design-with} humans and non-humans, rather than to \emph{design-for} an idealised `user'.

Posthumanist design can take many forms, including the \emph{counterfactual artefact} \cite{Wakkary_Odom_Hauser_Hertz_Lin_2015} (also common in Speculative Design \cite{Dunne2013}), a non-normative approach used to question design or technological conventions; \emph{slow technology} such as found in the work of Odom and colleagues, whose projects such as \emph{Photobox} \cite{Odom_2012,Odom2014}, \emph{Chronoscope} \cite{Chen_2019} and \emph{Olly} \cite{Odom_2019} explore time and memory using technology to envisage longer-term relationships with personal data and memories. With a similar approach to time, \emph{Long-living chair}---a rocking chair designed by Larissa Pschetz---records and displays its use over a period of ninety-six years, allowing the owner to observe long-term patterns of use \cite{Pschetz2013}. 

\changed{What distinguishes the research described here from this previous work in posthuman design is the integration of artificial intelligence technologies into the assembly of agents within the system. AI brings new forms of autonomy and agency into play, and this agency has unique characteristics that differ from more simple technologies \cite{Colton:2020aa}, including the ability to learn, change and adapt to people and their environment. These generative capabilities make traditional understandings of function or behaviour inappropriate, as such qualities may change over time and can not be fully anticipated in advance by their designers.}

\section{Message Ritual}

In this section we present an overview of \MR---an integrated system that aims to assist in the re-framing of memory and identity through machine generated poetics. Appropriating a standard domestic object---a table lamp---the work augments this domestic technology with a machine listening capability and a backend AI system. The lamp is designed to have an on-going presence in the home and functions just like any conventional lamp. \changed{Unlike domestic ``smart devices'', such as voice assistants, the lamp is not intended as a ``smart assistant'', being unresponsive to spoken commands and making no acknowledgement of listening or understanding spoken words  (c.f.~\cite{Reddy2021}). While other forms of interaction, including voice interaction, were considered in the early conceptualisation stage, these were rejected as likely being confusing (what commands does/should it respond to?); over-emphasising the novelty of ``smart devices'' at the expense of more meaningful and ritualised entanglements with technology.}

We explore an alternative approach to traditional augmented memory systems that assist in supporting a stronger sense of self-identity. Rather than focus on the sensing and quantification of external experiences, we aim to construct meaningful, personal narratives through the use of machine-generated poetics. \emph{Message Ritual} is an ambient Artificial Intelligence (AAI) system devised to prototype these ideas. Using speech recognition the system listens in on personal conversations---drawing out key words and phrases of significance---using them to generate poetic narratives that are sent to the participant's mobile phone as SMS messages each morning.

Using the lamp's ability to listen to human speech, \MR  seeks to draw out significant elements of on-going domestic conversation and use them as the basis for synthesising poetry specific to each member of the household. The process of listening and synthesising operates over a 24 hour cycle, with poems delivered to each member of the house via a text message when they wake each morning. The longer participants engage with the lamp, the better it gets at identifying topics of significance in their conversations. We now describe the technical implementation of the work.

\begin{figure}
  \includegraphics[width=\columnwidth]{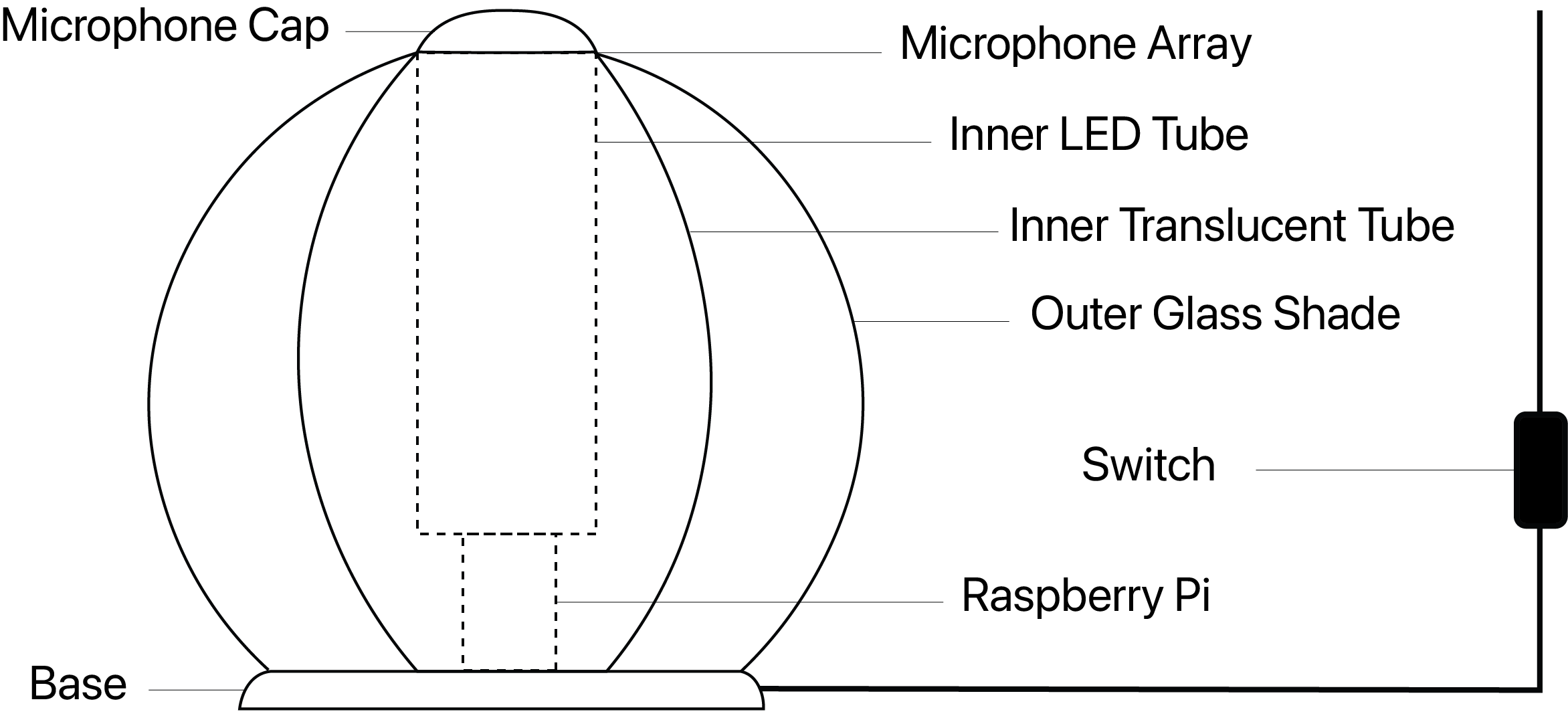}
  \caption{Components of the \emph{Message Ritual lamp:} a microphone array picks up nearby conversations when the lamp is switched on. On-board recording and speech detection is managed by a Raspberry Pi computer embedded in the lamp. A translucent inner tube with internal, programmable LED lights provides the lamp's light.}
  \Description{Components of the Message Ritual lamp: a microphone array picks up nearby conversations when the lamp is switched on. On-board recording and speech detection is managed by a Raspberry Pi computer embedded in the lamp. A translucent inner tube with internal LED lights provides the lamp's light.}
  \label{fig:lampDiagram}
\end{figure}

\subsection{System Architecture}
\begin{figure*}
  \includegraphics[width=0.9\textwidth]{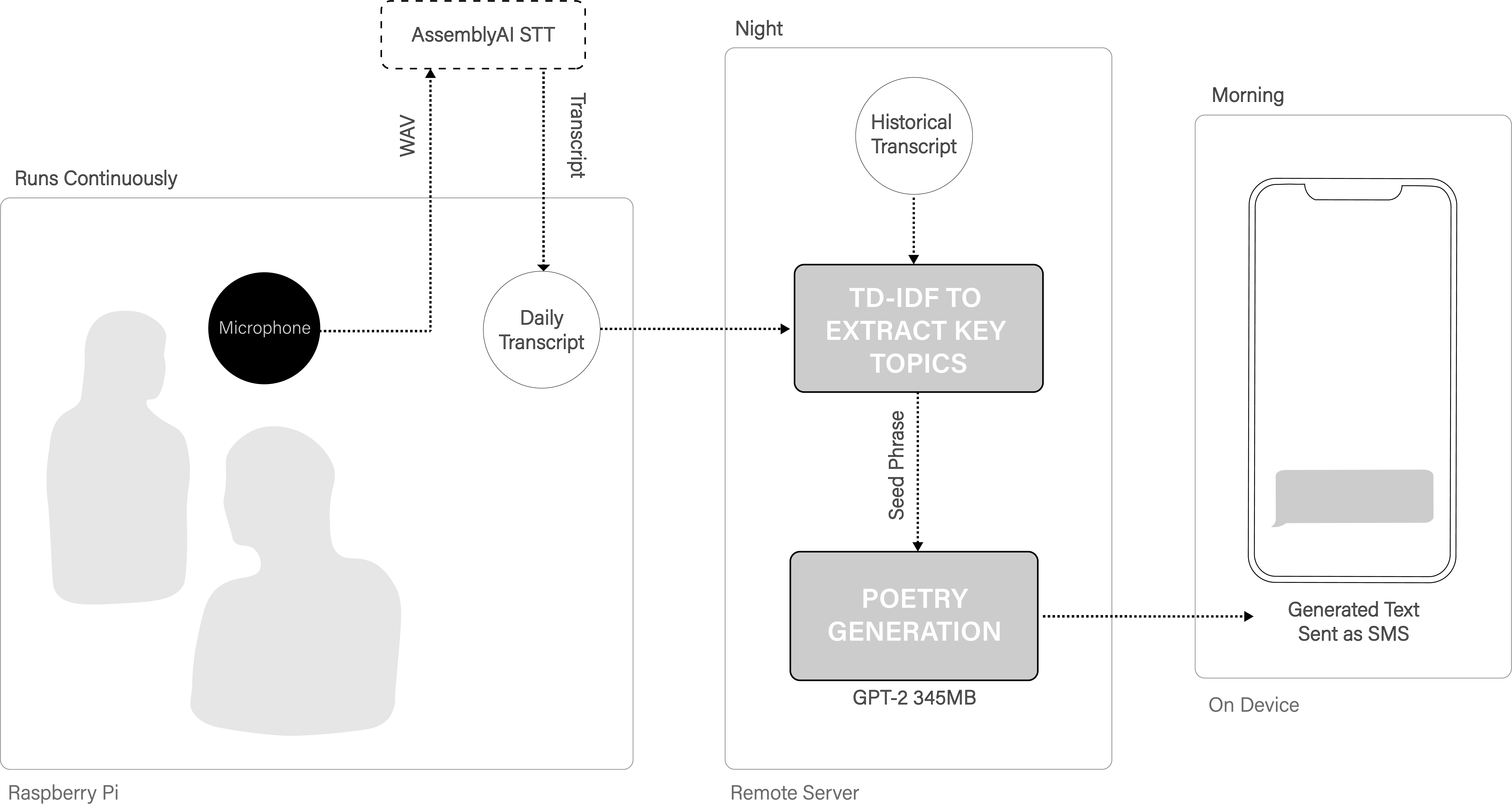}
  \caption{\emph{Message Ritual} schematic diagram. Conversations recorded by the lamp are sent to a cloud-based AI transcription service that returns text transcripts of the conversations. These transcripts are then sent to the \MR server, which analyses them to generate short poems based on the topics of conversation. Every morning a new poem is sent to each participant's smartphone as a text message.}
  \Description{Message Ritual schematic diagram. Conversations recorded by the lamp are sent to a cloud-based AI transcription service that returns text transcripts of the conversations. These transcripts are then sent to the Message Ritual server, which analyses them to generate short poems based on the topics of conversation. Every morning a new poem is sent to each participant's smartphone as a text message.}
  \label{fig:system}
\end{figure*}

The \MR system consists of three main components (Figure \ref{fig:system}): (i) a `lamp' with built-in microphones and computer, (ii) an AI speech-to-text transcription service; and (iii) a server application (\emph{Message Ritual Server}), which takes captured conversational transcripts and uses them to create custom poetry that is then sent to individual users as an SMS text message. We will now describe each component in more detail.

\subsubsection{Lamp}
\begin{figure}
  \includegraphics[width=0.9\columnwidth]{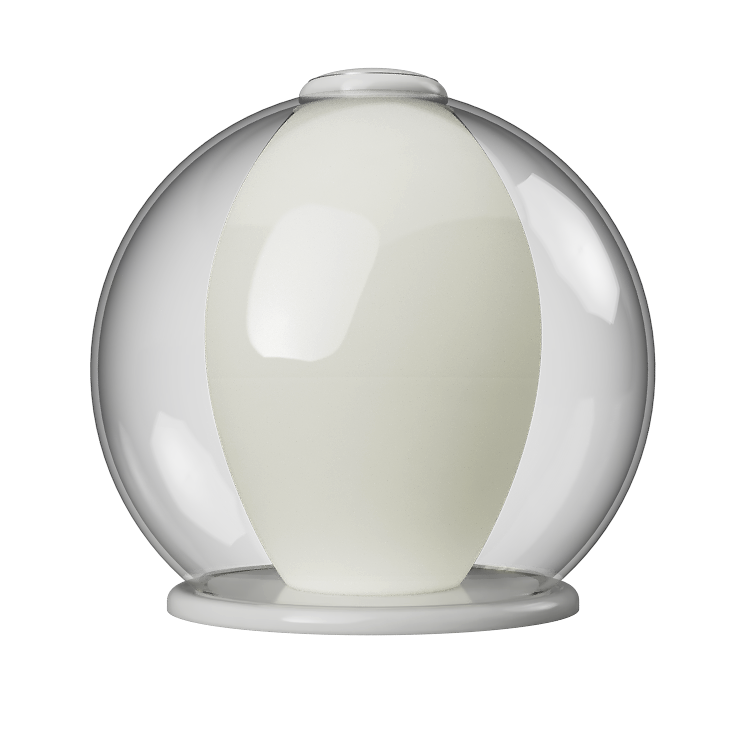}
  \caption{CAD rendering of the \emph{Message Ritual lamp}}
  \Description{A CAD (Computer Aided Design) rendering of the Message Ritual lamp, showing a glass sphere encasing an inner elliptical tube.}
  \label{fig:lampCAD}
\end{figure}

The lamp is designed with a warm, contemporary aesthetic to feature in a communal area of a home (Figure \ref{fig:lamp}). It consists of a translucent inner core with in-built LED illumination and an outer glass sphere, approximately 30cm in diameter (Figure \ref{fig:lampDiagram}). We used CAD design software and 3D printing to create this bespoke object (Figure \ref{fig:lampCAD}). The lamp has what appears to be a conventional power switch that actually serves two functions: to turn the illumination on and off (as would be the case with a conventional lamp) and to start and stop the listening process. When the lamp is illuminated it is listening for human voices, but when switched off the listening system is deactivated. This simple functionality gives participants an easy choice as to whether they would like their current conversations to form part of the \MR or not.

Inside the inner core of the lamp we embedded a microphone array based on the XMOS XVF-3000 DSP voice processor.\footnote{\url{https://www.xmos.ai/download/XVF3000-3100-product-brief(Sept-2020).pdf}} \changed{This processor is specifically designed for voice detection and transcription duties, automatically detecting the best microphone mix based on the direction of the detected voice and performing real-time processing on the incoming audio, such as echo cancellation, noise reduction and audio beam forming.} The digital audio output of the microphone array is fed into a Raspberry Pi SBC (RPi) housed at the base of the lamp. The RPi takes the digital audio stream and processes it through a WebRTC\footnote{\url{https://webrtc.org}} Voice Activity Detector (VAD). The VAD analyses each 30ms frame of incoming audio and looks for those frames that contain human speech. If speech is detected, the audio is captured and saved to a file. Listening and recording of conversations only occurs when the lamp is switched on and illuminated.

\subsubsection{Transcription}
The lamp captures these conversations throughout a 24 hour period and stores them as audio files locally on the Raspberry Pi, concealed in the base of the lamp. At regular intervals, the transcripts are uploaded to a speech-to-text transcription service. This service receives audio files, and, using a deep learning model, translates human speech into text. After evaluating a number of different speech-to-text services, we used the service offered by Assembly AI for the system described in this paper (\url{https://www.assemblyai.com}). Our evaluations were based on cost, ease of use (in terms of the programming API), accuracy in conversational settings and ability to deal with different English dialects and accents (as participants in our study spoke with different accents). While our system currently only operates with English speech, there is no technical or conceptual reason why it could not work with any major human language.

Once the recorded audio is converted to text by the transcription service, the text is returned to the Raspberry Pi in the lamp as json files that are stored locally over a 24-hour period. Once successfully transcribed to text, the recorded audio is deleted. At each 24-hour cycle, the Raspberry Pi uploads all the text transcripts received in the last 24 hour period to the \emph{Message Ritual Server}.

\subsubsection{Message Ritual Server}
Transcripts received from an individual lamp for the previous 24 hours are then processed to generate a unique poem for each member of the household currently using the lamp. For each active lamp, the server keeps a historical record of all conversations (the historical transcript) and uses this as a basis to extract the key elements and topics of each conversation. After filtering stop words, we used term frequency–inverse document frequency (TF-IDF, \cite{Salton1988}) analysis to select potential candidates as key words of each conversation. TF-IDF
analysis returns a weighted frequency that signifies the frequency of a given word in the previous day's conversation, as compared to it's frequency in the historical transcript. Thus, the longer people use the lamp to track their conversations, the better it becomes at evaluating potentially important topics of conversation. TF-IDF does this by recognising words that are frequently used in a new conversation but historically infrequent across all conversations. 

\subsubsection{Poetry Generation}
The top 20 words are then determined according to their weighted frequencies, along with their part-of-speech (POS) tags, with these randomly selected for each participant to form a \emph{seed phrase} for the poetry generation. Seed phrases are short concept ``seeds'' used by the AI poetry generation system to synthesise a unique poem. The general structure of the seed phrase is ``Adjective Noun, Noun''.
Our poetry generation system is based on the GPT-2-345M transformer network developed and released by OpenAI \cite{radford2019language2}.
The Transformer architecture was first introduced in \cite{vaswani2017attention}, and has since become a popular method to produce state of the art results in Natural Language Processing (NLP) tasks \cite{DBLP:journals/corr/abs-1810-04805, radford2019language2}.
We fine-tuned the network parameters of the 345M model by retraining it on a custom corpus of selected texts. The choice of training material directly reflects our overall design intentions: to encourage the reframing of personal memories through poetry. Texts were selected from a large range of styles and formats, including contemporary poetry, horoscopes, and philosophical aphorisms and writings. Texts were loosely chosen to satisfy one of two requirements. Firstly, we intend for the generated text to address the user directly by using the second-person pronouns `you' and `your'. This technique invites the user to make sense of the text through reflection on themselves and their experiences. Users will naturally place themselves as the subject of the lamp's poetry, and through continued use should develop a relationship with the system. Secondly, the content of the texts selected should be of a personal nature, emotionally charged or contain strong imagery. 

The final corpus is composed of 2,605 individual short-form texts, reaching just over 1MB in size. At this size, overfitting our model becomes a concern. We do not want the lamp to simply replicate full poems from our corpus, but to produce original output that blends the various writing styles. 
The text generation process takes as input a number of parameters: \textit{length} is set to 120 (upper limit on word length of output); \textit{temperature} set to 0.9 (higher temperature results in more random completions), and \textit{top\_k} set to 80 (80 words are considered at each step of generation). The \emph{seed phrase} is used to initiate the generation of a poem. The output text is trimmed to prevent run-off sentences, and furthermore discarded (and regenerated) if the length does not fall between 50 and 450 characters. We found this range of lengths is most suitable for the SMS format. Separate poems are generated for each member of the household in which the lamp is located and sent by SMS in the early morning.

\begin{figure*}
  \includegraphics[width=0.49\textwidth]{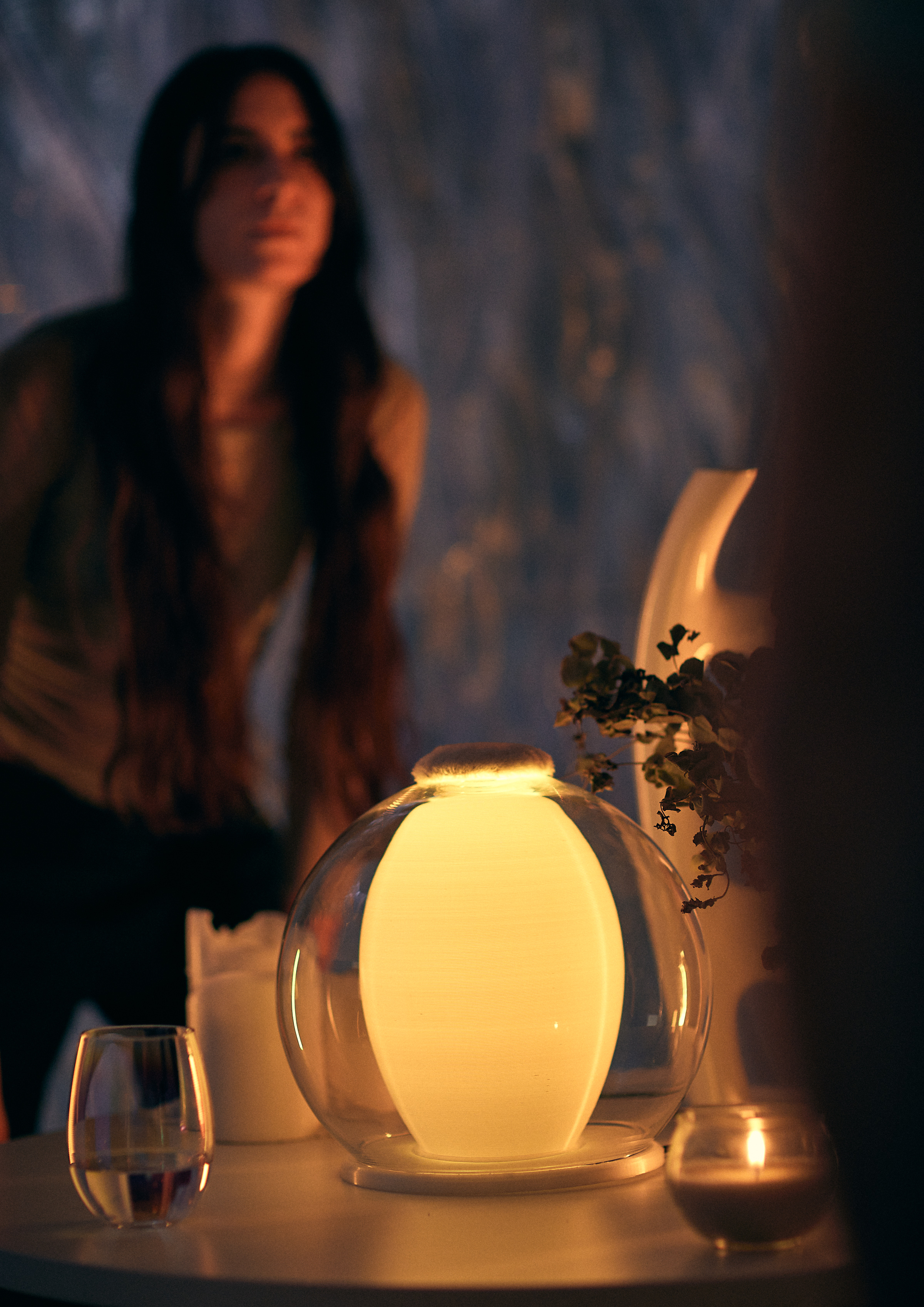}
  \includegraphics[width=0.49\textwidth]{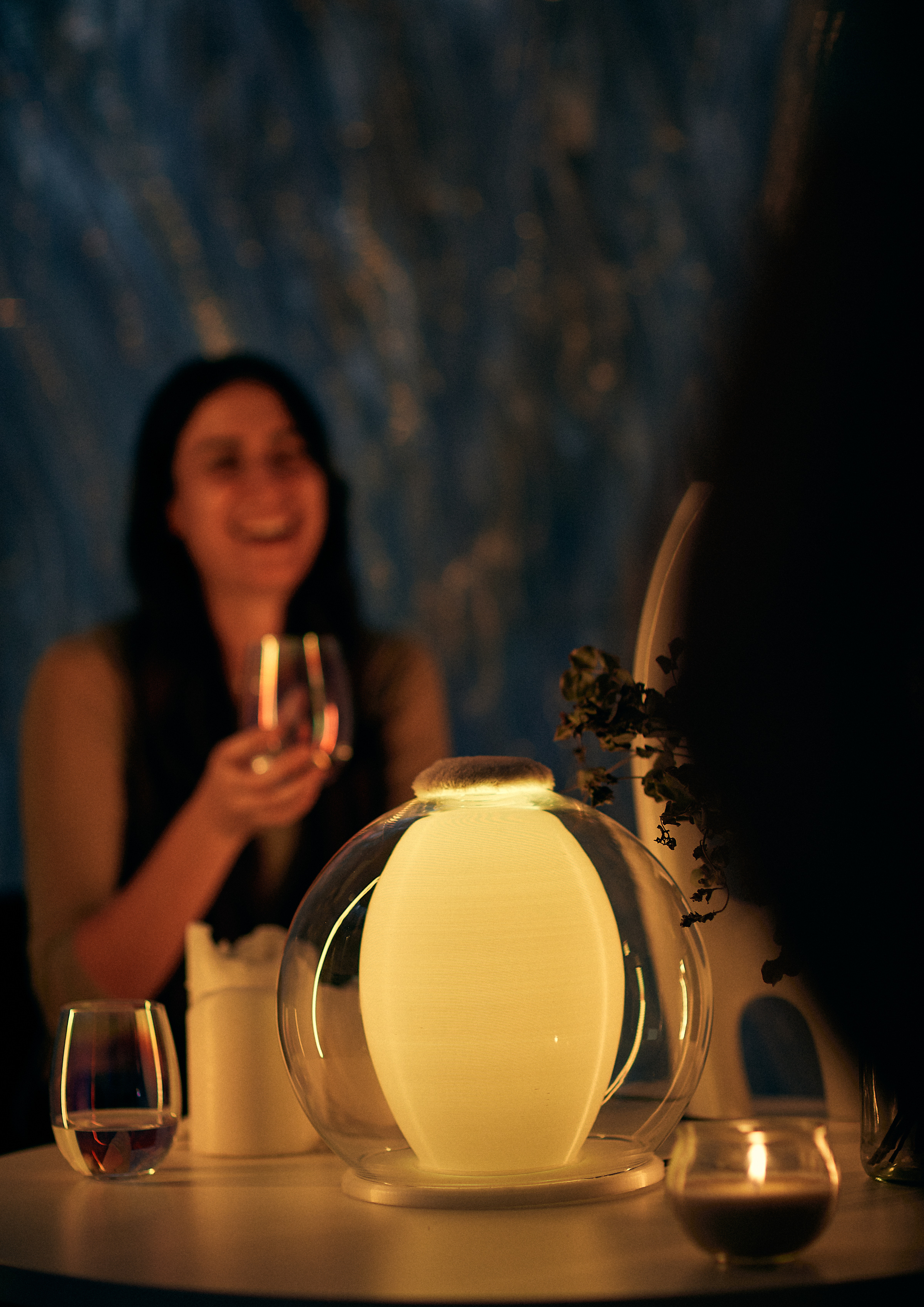}
  \caption{The \emph{Message Ritual} lamp in a domestic setting. The lamp blends into the space, its light carving a space for people to gather and converse.}
  \Description{The Message Ritual lamp in a domestic setting. The lamp blends into the space, its light carving a space for people to gather and converse.}
  \label{fig:lampConversations}
\end{figure*}

\subsection{Design Factors}
This work draws on the development of \emph{Mirror Ritual} \cite{Rajcic_2020}; a real-time affective interface in which machine-generated poetry is dynamically presented as a response to perceived emotion of the viewer. \MR similarly utilises machine-generated poetry to foster reflection and introspection as the individual attempts to draw meaning from their delivered poem. Instead of grounding itself in the viewer's in-the-moment mood, \MR draws upon an individual's recent memories to contextualise their poetry. The system is not attempting to improve the recollection of `accurate' memories, rather it encourages the active framing and re-framing of personal events that contribute to one's overall life narrative. The work is designed to be lived with over an extended period of time, with the poetry referencing the events of the previous day. This self-referential nature of the experience aims to provide some sense of continuity over the interactions, allowing people to develop a meaningful relationship with the system over time.

The design of the lamp itself is crucial to the overall experience, as it will be embedded into the homes and lives of it's users. Similar to \emph{Mirror Ritual} \cite{Rajcic_2020, Rajcic_TEI_2020}, the lamp serves both a functional and aesthetic role in participants' homes. Everyday domestic objects (e.g. furniture, crockery, plants, soft furnishings) are of interest here as they have clear and well-defined roles in one's life, but they also distinctly lack a technological component. In wider HCI research, we have seen the widespread integration of technology and everyday objects (the `smartification' of the home) \cite{bello2019toward, Rozendaal2021}, generally aiming to augment both the function of the artefact itself, and to increase the connectedness of objects with each other (i.e.~internet of things). In this project, an everyday object is chosen not to necessarily `improve' or augment that object's function, but rather to bypass any preconceived notions or behaviours that users may associate with traditional interfaces such as screens and smartphones. In this way, the natural behaviours afforded by non-technological objects (i.e. switching a lamp on or off) are utilized in the interaction, allowing for the interface to escape common behavioural and gestural habits such as tapping, swiping, scrolling, refreshing etc.

The artefact blends into domestic surroundings (Figure \ref{fig:lampConversations}), so as to not interfere with natural conversation occurring in the household. Furthermore, the light from the lamp carves out a space for occupants to gather and converse, especially in the evening and night time hours. 

A smartphone was chosen to serve as an intervention to the typical habitual behaviours and gestures associated with the device. The morning SMS is designed to reach the participant in their first moments of the day, encouraging reflection and re-framing of the previous day's events. The daily reading and interpreting of the message contributes towards one's waking ritual, prompting the participant to proactively reflect and contemplate their life. SMS technology is utilized as it sits outside of social media applications, creating a conceptual division between \MR, and other morning routines that are mediated through a smartphone.

\begin{figure*}
  \includegraphics[width=\textwidth]{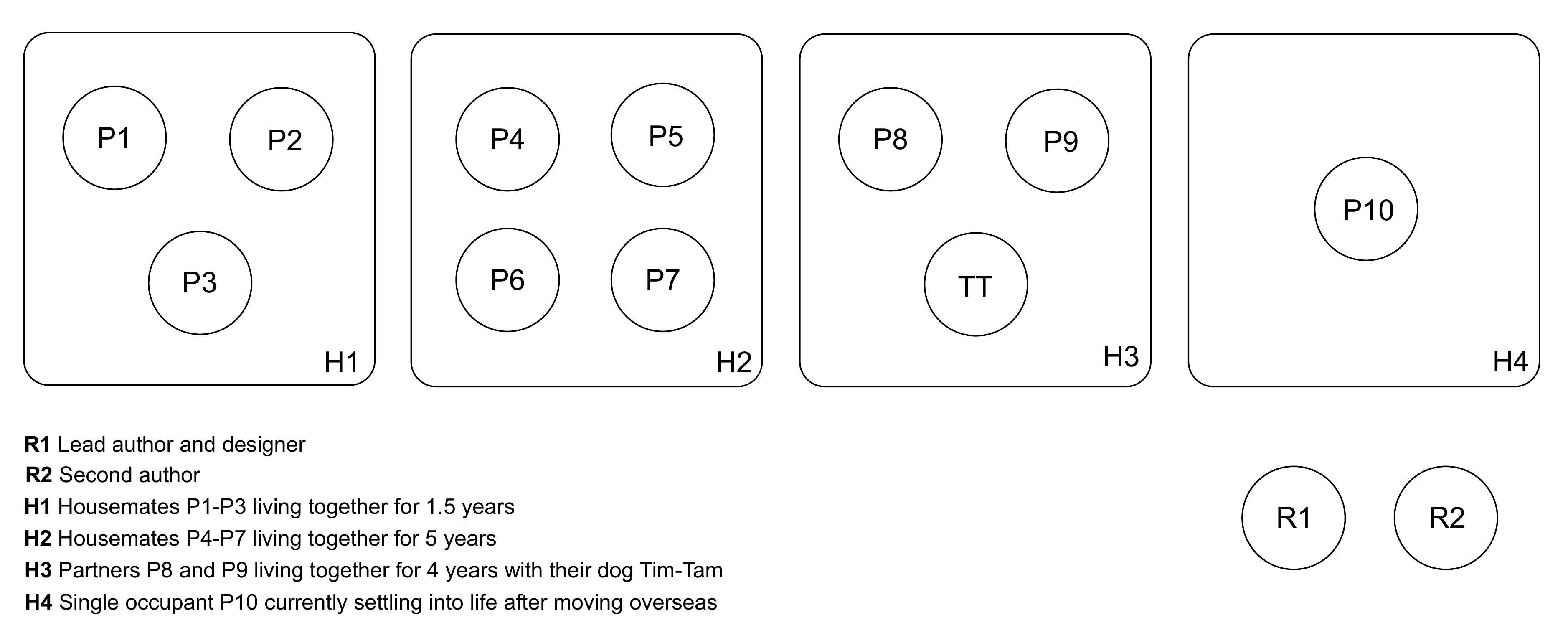}
  \caption{Diagram illustrating the configuration of each household, including information on the pre-existing relationships between participants and researchers}
  \Description{Diagram illustrating the configuration of each household, including information on the pre-existing relationships between participants and researchers}
  \label{fig:configuration}
\end{figure*}

\section{Evaluation}

Our approach to evaluation of \MR is inspired by Barad's diffractive methodology, in turn inspired by feminist writing of Haraway \cite{haraway1992promises}, and proposed by Frauenberger as an epistemological framework in HCI \cite{frauenberger2019entanglement}. Rather than striving to determine whether an artefact works at achieving a hypothesised outcome, we instead strive to understand what an artefact becomes as it is put out into the world. This is done by not only recognising \emph{sameness} across data (e.g. via thematic coding), but through identification and exploration of the \emph{differences}. The application of diffractive methodologies in HCI has received increased interest recent years. For example, Lazar et al. apply three distinct analyses to the same set of qualitative data gathered from workshops exploring ``technology for aging''. The results from each technique were then read \emph{through} each other to better understand how the apparatus (the empirical analyses in this case) co-constitutes the phenomena being observed \cite{10.1145/3462326}. In other case studies, diffraction has been used as a framework through which to approach artistic practice \cite{scurto2021prototyping, nordmoen2022making}. The diversity in approaches to diffractive analysis demonstrates that it does not encompass a fixed methodology, but rather an open reading of data `through' or alongside other materials such as ``texts, personal experiences, other data'' \cite{doi:10.1177/13607804211029978}. 

\MR was not designed with any specific purpose in a given household, but rather to explore the open-ended possibilities of living with a non-human system with its own agency and autonomy. Taken into consideration during the conception and design process were general themes that relate to the material and functional affordances of the system, which include conversation, language, memory, ritual, domesticity and of course, light. In the following evaluation, we adopt a diffractive methodology in the analysing interview data gathered across four households living with the lamp. In particular, we are interested in how the lamp \emph{becomes} different things when entering into assemblage with differing household configurations. For this reason we have selected households that span a range of domestic living situations, from single member households to shared living arrangements.

\subsection{Study Design}
To study the system's entanglements with people, 10 participants (6 male, 4 female) across four households were recruited to live with the lamp over a two week period. Recruitment was from members of our University and through personal contacts. \changed{7 of the 10 participants were known to the researchers personally, but none were familiar with the work or its theoretical foundations at the commencement of the study.} No remuneration was provided for participation. Participants' ages ranged from 27 -- 36 with a median of 31. \changed{We acknowledge this is a relatively narrow age range, determined largely by the demographics of participants who volunteered to be involved in the study.} Participants were asked to naturally incorporate \MR into their daily routine, turning the lamp on when engaging in conversation, and off during quiet hours, or when they required privacy. Each morning, the participants received a unique poem via SMS at the approximate average wake time of the household. Prior to the study, participants were made aware of the basic functioning of the lamp -- namely that speech is recorded on the device and then transcribed, with the resulting transcripts being sent to a remote server for analysis and poetry generation. Participants were comprehensively briefed on any privacy concerns, including the treatment of audio files, and the privacy policies of third party providers. The study received full ethics approval from our University's ethics panel prior to commencing.

The details of each household configuration can be seen in Figure \ref{fig:configuration}. The households were selected to cover a variety of living arrangements. Households 1 (H1) and 2 (H2) both comprise of shared living situation between friends, with 3 and 4 household members respectively. The primary difference between H1 and H2 is the length of relationship, with members of H2 having lived together for over 5 years, where H1 is comprised of relatively new relationships. Household 3 (H3) comprises of a couple who have been living together for 4 years, with their dog Tim-Tam (TT). Finally Household 4 (H4) contains a single member P10, who has recently moved country and is settling into a new living situation.

At the conclusion of the study, participants took part in a semi-structured interview regarding their overall experience living with lamp. Their responses were thematically coded to identify common themes and experiences across participants. The interview data was again read `through' these emergent themes, along with relevant theory, researchers' personal experiences, and experiences of participants.
In the following diffractive analysis, we identify and unpack the shared and the distinct experiences across both individual participants and collective households, aiming to explore what the lamp \emph{becomes} as it enters into assemblage with households and researchers.

The study is exploratory in nature, designed in order to gather key insights into the design considerations of the system at the initial stages of its entanglement with human participants. As a consequence, the study has a limited scope in both number of households, and length of participation.
Given these limitations, we nevertheless feel that such exploratory investigation is crucial in the building of personal and domestic technologies, and ultimately calls for critical reflection and ethical accountability at the early stages of development.
We intend for the insights gained in this exploratory study to inform the subsequent design iterations of \MR, as well as future evaluation methodologies.

\subsection{Habit and Ritual}

Parts of the study were coincidentally held during periods of `lockdown' as a result of the COVID-19 pandemic.
Several participants suggested that their experiences with the lamp were intrinsically tied to their experience of increased social distancing and isolation. One participant noted that in lockdown, it's difficult to ``differentiate between days'' (P1, P6), and so the daily ritual of conversing, reading messages, and reflecting offered a ``point of difference'' (P1) against the monotonous backdrop of social isolation. All participants valued the addition of \MR to their regular routines, which was most notably demonstrated in the disappointment experienced on the occasion that the lamp failed to send a message (due to insufficient conversation). A majority of participants used the word ``disappointment'' to describe how they felt even though, when prompted, they found it difficult to explain what exactly they missed.

Viewed through Han's conceptualisation of ritual as the furnishing of time, we can see that \MR offers structured time in which participants are allowed to linger. The poetry is not consumed, but rather it ``endures'' \cite{Han2020}, sometimes opaque and difficult to parse, participants described dwelling on their morning message from anywhere between 30 seconds to 30 minutes. Similarly, the time required to produce sufficient conversation is not trivial, and participants found themselves carving out time for more in-depth conversation centred around the lamp, with H1 structuring discussions around meal times, away from TV and other distractions.

\begin{quote}
P2: ``We met more often at the dining table to talk, whereas usually we would just talk in passing in the house, it kind of gave us a central point to have those conversations\ldots it's been really nice''
\end{quote}

Similarly, participants of H3 found more time was spent conversing around the dinner table (where the lamp was situated) so as to ``give it more conversation'' (P8). In asking something of participants (sufficient conversation, the reading of poetry), the lamp in return offered regular and reliable moments of structured time, that largely came as respite from the overabundance of time that lockdown affords.

P2 and P3 shared their recent efforts of reducing the amount of time they spent on their smartphones, citing social distancing as the reason for falling into undesirable habits. Both participants describe the ``losing'' (P2) or ``destroying'' (P3) of time that occurs when on their device, both illustrating the material interventions required when sheer willpower was not enough. From leaving the phone in a different room, to the adoption of a second ``boring'' (P2) phone that is used exclusively throughout the working week. Although not explicitly referencing the lamp, such conversations further contrasted the types of agencies at play between human and machine. Returning to DeJong's account of the `Feed' \cite{dejong2019posthuman}, we here see the trace of non-human agencies and their very real and material effects. Both P2 and P3 expressed that their engagement with \MR did not have any effect (positive or negative) on this desire to limit or disrupt phone use, and that these material interventions were a necessary measure to exert their own agency in their relationships with the `Feed'.

\subsection{Memory and Identity}

\begin{figure}
  \begin{center}
    \includegraphics[width=0.9\columnwidth]{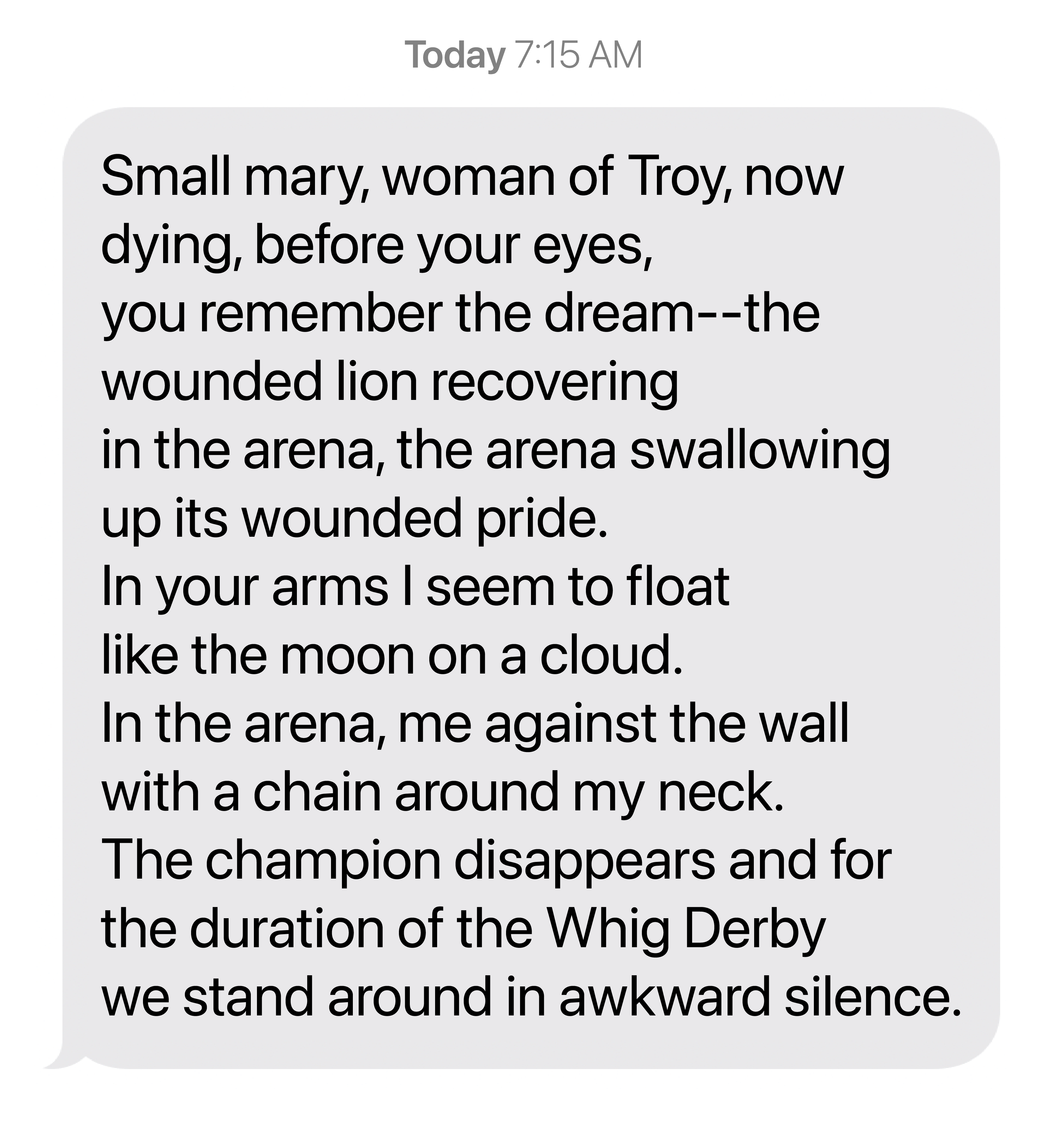}
  \end{center}
  \caption{Message received by P7}
  \label{fig:msgp7}
\end{figure}

The interpreting of poetry came naturally to all participants, who were each at times able to meaningfully connect the poetry back to conversations from the previous day. At other times, participants saw no obvious connection, and instead were reminded of moments from childhood (P1), of friends and family they hadn't seen in a while (P4, P5), of past holidays in Rome (P7) (Figure \ref{fig:msgp7}), of books they had read (P3). A number of participants viewed the messages as the lens through which to remember previous conversations, describing them as a ``puzzle'' (P6) to decode, met with a ``sense of satisfication'' (P2) once having solved it.
Others struggled with the uncertainty of their own memories, reflecting on whether the lamp's memory was sometimes more accurate than their own.
\begin{quote}
     P4: ``It’s like the lamp is reflecting back memories through its own lens, right, which is different to mine. Would you say it’s an objective lens, or has its own bias?''\\
     Interviewer: ``What do you think?''\\
     P4: ``It feels more objective but maybe that's not the case. To me it reflects what is. It reflects what we put into it.''
\end{quote}

The tendency to perceive the machine as being altogether more objective, rational, or accurate than the human has roots in early computing culture \cite{10.1145/255950.153587}.
Barad's account of the role of measuring apparatus provides an understanding of the lamp as that which records memories. 
In the way that `the apparatus is an inseparable part of the observed phenomenon.' \cite{barad2014diffracting}, so too is the lamp itself inseparable from the memories constructed and reconstructed through its intra-action with participants. In this way, the phenomenon of memory recollection is performed with and through one's relationship with the lamp, other house members, and researchers within an assemblage. P6 describes how the lamp's language was entangled with their own language, and with the language of other house members, to the extent that it became impossible to differentiate or trace back to a single source. Instead, remembering through the imagery and narratives in the poetry represented a kind of collective memory within the household. 

For other participants, the lamp's capacity to provoke recollection had less to do with the specific memories surfacing, and more to do with the very act of remembering. In this way, the lamp served as a ``reminder'' (P2, P6) to remember, which at times led participants on a path of reflection, introspection and self-awareness at the outset of their day. Several participants recited the cascade of thoughts that followed the reading of a poem. In one example, a poem urging P7 to be ``open and honest'' with themselves led to them contemplating their feelings towards a recent ex-partner. The lamp provided further ``affirmation'' (P7) of existing feelings centred around the decision to end the relationship earlier in the week. 

In \cite{wilson2003identity}, the authors propose that the primary function of Autobiographical Memory (AM) is to form a coherent sense of one's identity. Personal memories are said to shape one's current sense of self, and in turn this self-identity influences how past experiences are remembered. In a similar vein, \cite{conway2005memory} proposes that autobiographical memories are dynamically retrieved and reconstructed to serve the current goals, beliefs and image of the \emph{working self}.
Such approaches to understanding AM highlight its crucial role in the sense-making of the present: in order for the present self to remain coherent, the past must be perpetually malleable. This organisation of autobiographical memories to construct a consistent, continuous subjective experience can be likened to the authoring of an autobiographical narrative -- a narrative that is continuously subject to revision.
We found that for most participants, the value of the lamp is not in its ability to produce an accurate summary of the previous day's conversation, but rather in its proclivity to encourage such authoring -- often giving participants the language through which to frame their recent experiences.

\begin{figure*}
  \includegraphics[width=\textwidth]{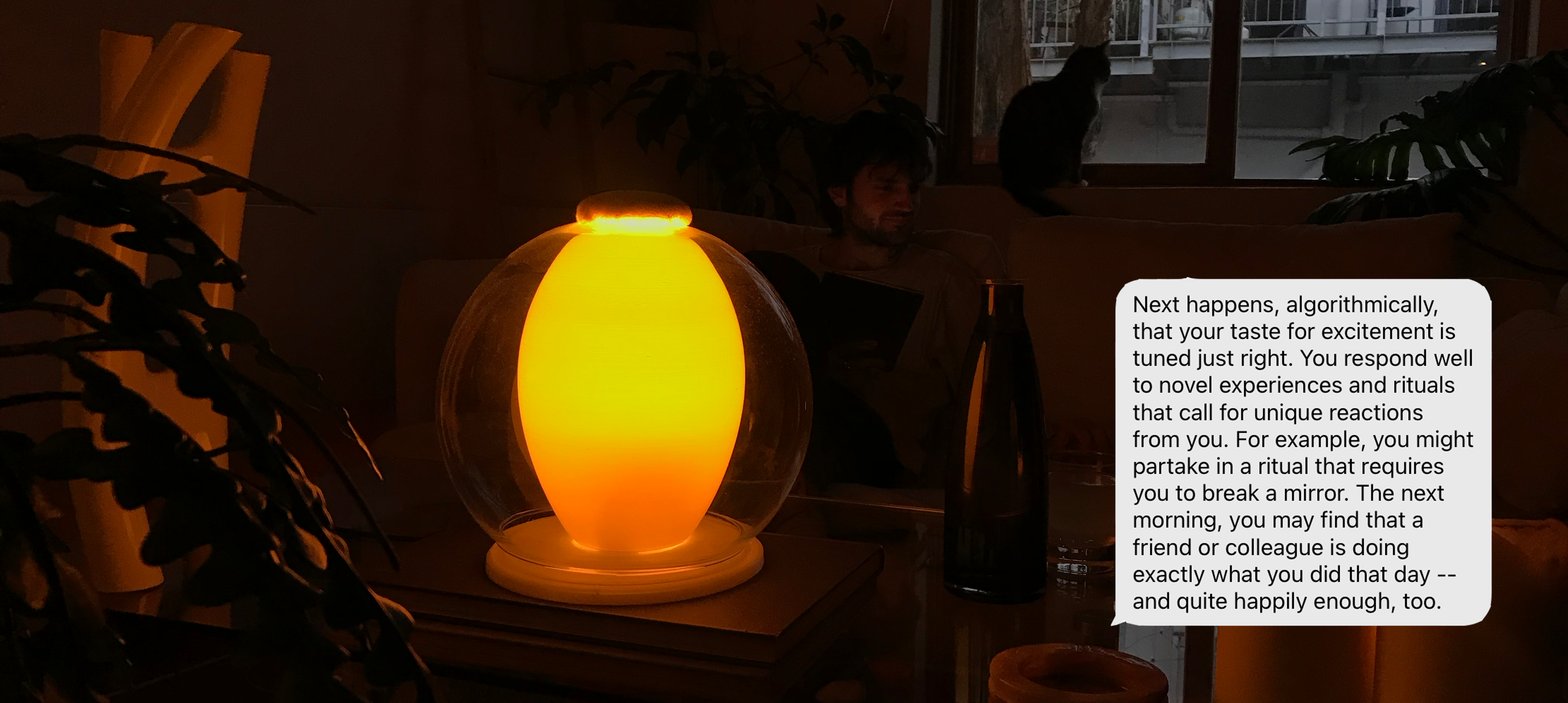}
  \caption{Lamp pictured alongside message received by P1 on the first day of the study}
  \Description{Lamp pictured with message received by participant overlaid}
  \label{fig:greeting}
\end{figure*}

\subsection{Relationships}

The majority of participants found the lamp to be a good ``conversation starter'' (P7), with households H1-H3 regularly sharing and discussing their respective poems. Along with being given a ``reason'' (P6) to converse more regularly, participants now too had ``something extra to talk about'' (P2). After a while, this process became ``cyclical'' (P5), with the poetry leading ``into more conversation for the lamp for the next day'' (P5). This shared experience fostered a ``sense of togetherness'' (P7) across the larger households (H1, H2). Returning again to Han's account of ritual -- it is not just that the rituals allow for lingering in time, but they also entail ``community'' \cite[p.43]{Han2020}.

Participants of H3 similarly describe the lamp as a social agent, however not necessarily fostering relationships between household members, but rather the relationships with house guests. The lamp presented an added novelty, an ``extra guest at [their] dinner parties'' (P8). P8 and P9 reported that their guests would alter their behaviour around the lamp; at times they would check that the lamp had caught what they said, at other times guests would speak directly to the lamp in an effort to influence the next morning's poetry.
\begin{quote}
P8: ``Maybe it worked more as a social tool for us in that we enjoyed most when we had people over, and we really enjoyed sharing the poem, and then we would have a conversation with that group the next day''\\
P9: ``People were excited about it the next day, they’d ask `Did you get the poem yet?'''
\end{quote}

It takes two (or more) people to make a conversation, raising the question of how a person living alone might engage with \MR. In contrast to the relatively social households H1-H3, H4 comprises of a single participant (P10) who has recently moved country and is settling in to a new living situation. In the absence of other people, P10 attempted to include the lamp in external communication, like speaking ``on the phone'', listening to ``voice messages and podcasts'' (P10). P10 also describes their efforts to talk directly to the lamp: ``I was speaking to myself as well, I normally don't do that intentionally, but the past few weeks I've been like, okay, let it out, just talk to yourself'' (P10).

Language is often placed in opposition to the material under representationalism, which adopts a binary of words and things; signifier and signified. Yet, as Clark describes, language itself has a materiality; we encounter `words in the air, symbols on a printed page' \cite{clark1998magic}. Language is not merely a vehicle through which we express our inner thoughts, but a form of computation in itself. The supra-communicative view of language, originally pioneered by \citet{vygotsky1962thought}, proposes that language is a tool that guides behaviour and structures action. For P10, spoken language was not solely used as a method of communication, but as a tool to regulate mood, or structure thought. P10 describes the lamp as a sort of spoken diary:

\begin{quote}
    P10: ``Normally I don't practice that daily. But yes, I have for the past few weeks, like I said, it's refreshingly easy. I write diaries and I've done that probably 20 years now, Almost daily, so for me it was just kind of a nice to see it in an object instead of my actual physical diary...I was able to be vulnerable with it.''
\end{quote}

What role did the lamp take in the existing households? At first, participants describe an uneasiness with the introduction of the lamp into their space. Unsure of how the lamp would utilise their language, participants were markedly more ``self-conscious'' (P1) and self-censoring at the beginning of the study.
\begin{quote}
P3: `` In the early days there probably was less talking around the lamp, and then maybe towards end that sort of dropped away\ldots''\\
Interviewer: "What do you think changed?"\\
P3: ``Probably because the poems kept coming in and y'know, [I was] having fun with them''
\end{quote}
As the poems began arriving (Figure \ref{fig:greeting}), and familiarity grew, all participants started to become comfortable in its presence. The relationship developed towards the lamp itself varied across households, with participants describing it as a ``companianable'', ``a good listener'' (P7), and the new ``housemate'' (P2, P4, P5); ``it made me feel like I have the floor'' (P5). P7 described feeling flattered, like they were the lamp's ``muse'' (P7). Other participants describe moments in which they speak directly to the lamp, for example greeting it in the morning or asking it questions directly (P1). A majority of participants found themselves contemplating the thoughts or mood of the lamp, ascribing to it some level of intention, autonomy and agency; the lamp is ``not a human being, but something, it’s doing something'' (P7).

Perceptions of agency and autonomy are increasingly important in this new era of human-machine relationships, particularly when those relationships involve AI. The traditional anthropomorphisms of `smart devices', such as disembodied voices, skewmorphisms, or human iconography serve to reinforce the notion that artificial intelligence is akin to human intelligence. This sets up expectations of intentionality and free-will that we associate with the `theory of mind' type of autonomy \cite[Ch.9]{boden2012}, rather than other, potentially new forms of agency and autonomy possible in advanced technological systems.

\subsection{Transparency}
\label{ss:transparency}
\begin{figure}
  \begin{center}
    \includegraphics[width=0.9\columnwidth]{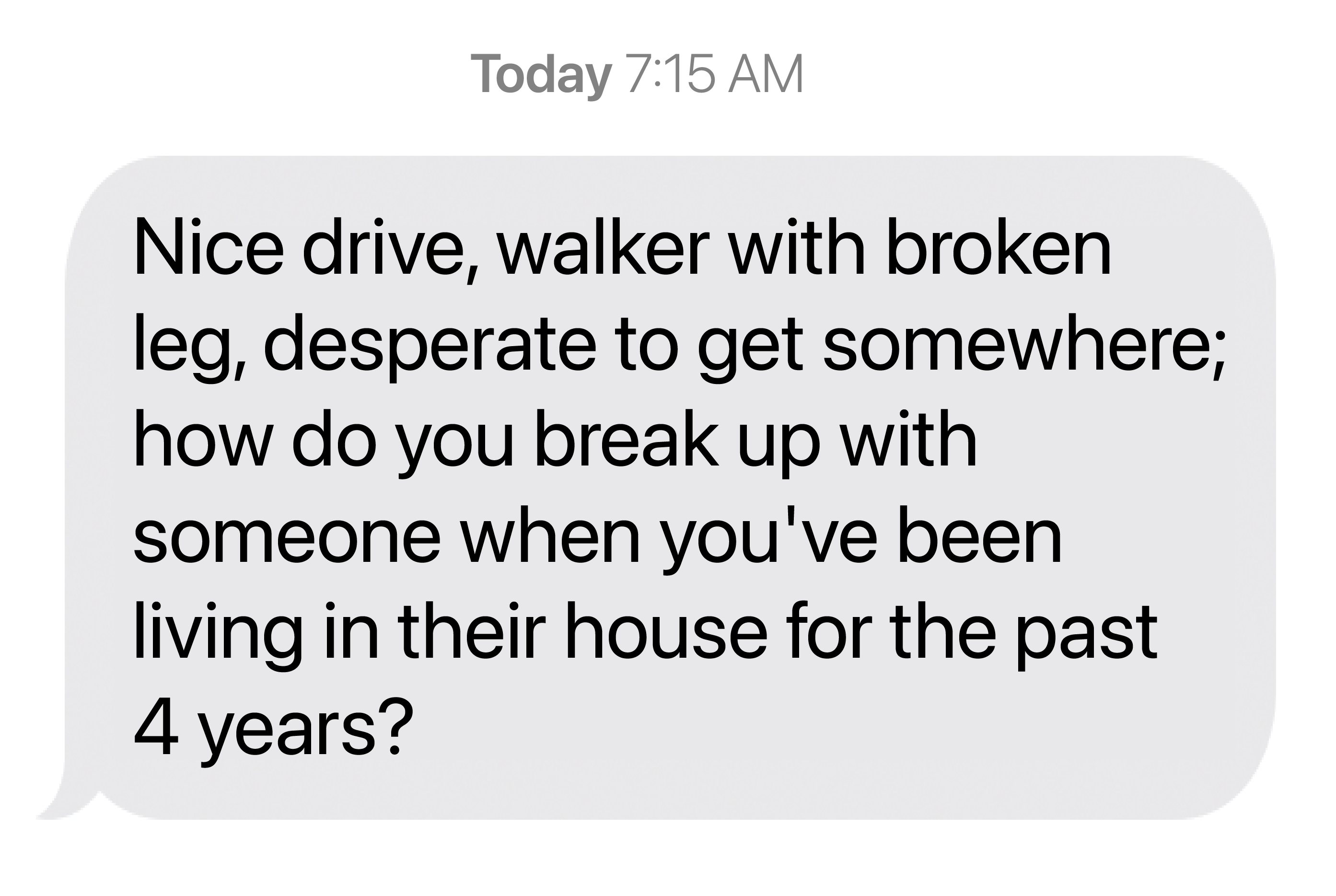}
  \end{center}
  \caption{Message received by P8}
  \label{fig:msgp8}
\end{figure}

The participants each had varying levels of knowledge and experience with AI technologies, with some participants directly involved with the development and design of similar systems (P2, P9), and others claiming almost no understanding of AI (P1, P8, P10). Gaps in working knowledge of the capabilities of AI technologies led to participants often over or underestimating the functionality of the lamp. In reference to a particular poem with the word ``barking'', P9 asked whether the lamp was able to correctly tag and interpret their dog's barking from the previous day. P9 also considered that perhaps they instead might have used that word in conversation.
In instances where the lamp's messages were unexpectedly dark and melancholy, P8 began to wonder whether the lamp was ``picking up on something'' (P8) beyond merely their language, like the tone of voice, or the ``vibe'' (P8) of the room. Rather than immediately reject the poetry that they could not clearly associate to the previous day, P8 instead read further into the lamp's capacity for `knowing'. 
\changed{In another instance, P8 recalls receiving a message (Figure \ref{fig:msgp8}) that contained an uncanny coincidence.}
\changed{P8 and P9 had in fact been living together for 4 years, yet they quickly rejected the perceived tone and intention of the message. What was ultimately written off as coincidence, nevertheless generated an emotional reaction, and led P8 to question where exactly lies accountability of such generative systems; stating that ``this would really suck if you were the one that wanted to [break up]'' (P8)}

The lack of transparency surrounding generative AI systems contributes to both its creative potential (or perhaps, illusion), and to the ethically challenging aspects of its application in domestic and personal contexts. On one hand, this ambiguity lends itself to different expectations: the opening up of new possibilities and alternative perspectives. However this same ambiguity may also lead to confronting, negative, or disturbing experiences. In our case, increasing transparency of the system (i.e.~explicitly reiterating data collection and storage procedures) relieved much of the concern for participants around privacy (P1, P3). Revealing the underlying transcripts to participants (P8, P9) of H3 furthermore brought clarity to their questions around perceived epistemic capabilities of the lamp, and perhaps highlighted the extent to which these observations, memories, and feelings were co-constructed through their entanglement.

\subsection{Entanglements}
\label{ss:entanglements}
P1, P3, and P8 were not familiar with previous research nor with the researchers at the beginning of the study, whereas all other participants were considered colleagues or personal contacts. It quickly became evident how the entanglement of the researchers in this assemblage influenced the perceptions and attitudes towards the lamp. \changed{In contrast with traditional analyses, our diffractive approach allows us to unpack the consequence of this difference, acknowledging that such research can not be not carried out in a vacuum.} 

P1 and P3 both likened the lamp to the famously cursed late 90's children's toy, Furby \cite{Marsh2019}, speaking to the perceived unfamiliarity or alien-like quality of the lamp. P8 reported never personally switching the lamp on or off, leaving that task to their partner who had more familiarity with the system (P9).

At times, participants felt guilt towards the lamp when they failed to turn it on for a day (P2, P6), although it was unclear towards who exactly the guilt was directed:
\begin{quote}
    P2: ``It kind of felt like we failed, actually, we didn't do the task we were supposed to do...''\\
    Interviewer: ``You failed me? Or you failed the lamp? Or you failed yourself?''\\
    P2: ``Good question, um, I think we failed ourselves''
\end{quote}

Participants with whom the researchers had a closer relationship to (P2, P4-7, P10) instead presented with a tendency to perceive the lamp as friendly and ``warm'' (P4), being less concerned with issues of privacy and overall more receptive to its language.
\begin{quote}
P5: ``It was an extension of you''\\
P4: ``It was like a housemate''\\
P5: ``We also spoke about you to the lamp, if you came up we would address the lamp directly as 'your mother'''
\end{quote}
Similarly, while using the lamp as a spoken diary, P10 at the same time likened it to a kind of mediated communication with the researchers; ``I was trying to share my gossip with you, I was giving all the juicy details to you'' (P10). 

The lamp served not only as an extension of its designer, but led to the further entanglement of researchers with both households through regular check-ins. The lamp, although undeniably exerting its own agency, seemed to also borrow agency from the humans that created it; trust towards the lamp was an extension of the designers trust, guilt felt towards the lamp was guilt towards the researchers, and the `success' of the study. This borrowed agency and trust heavily skewed the perceptions of \MR causing participants to be more receptive and forgiving of the system. As raised by participants of H3, if the lamp was instead a ``consumer product'' (P8), their experiences would perhaps be more troubling. Emancipated from a sense of human accountability and care, the lamp's messages could be perceived with an entirely different tone or intention. Returning to the discussion on ethical issues and of increased transparency of AI systems, we see that this transparency does not concern only the technical processes at play, but the tracing of human involvement.

\changed{Ethical concerns around data-collection practices can typically be traced back to the (human-led) corporations that employ them for profit \cite{Zuboff2018}. As discussed in \cite{breidbach2020accountable}, in order to gather data, ``the true purpose of data-driven value propositions is often concealed, and individuals are coerced into using them. This constitutes exploitation and constrains the ability of individuals and organizations to act independently'' \cite{breidbach2020accountable}. Building on a comment from P8, if the lamp was indeed a consumer product, this would raise serious concerns around the true value proposition of the service, and whether or not the hyperpersonalised nature of the system would inevitably lead to a unethical implications.}

\begin{figure*}
  \includegraphics[width=\textwidth]{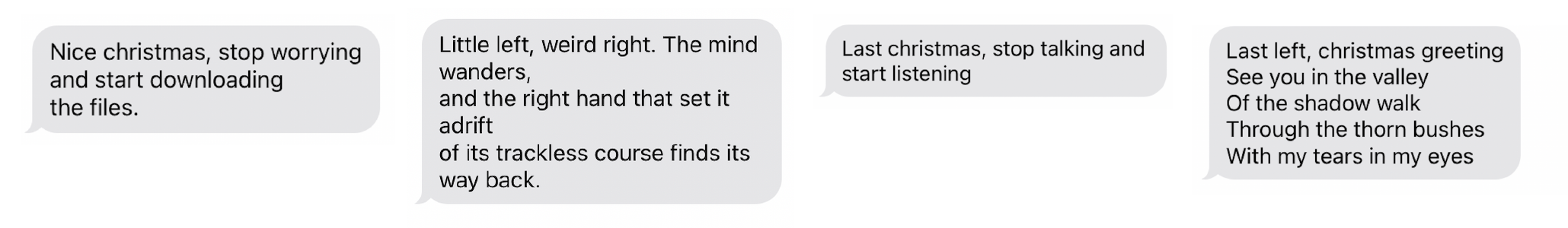}
  \caption{Goodbye messages received by participants of Household 2 on the last day of the study}
  \Description{Goodbye text messages received by participants of Household 2 on the last day of the study}
  \label{fig:goodbye}
\end{figure*}

On the final day of the study, participants of H2 each recount receiving what they interpreted as a ``goodbye message'' (P4-7) from the lamp (Figure \ref{fig:goodbye}). P4 expressed that ``it felt like the lamp is gonna miss me, it really created an emotion there'', speaking again to the participants proclivity to narrativize their interactions with the lamp. In contrast, participants of H3 instead experienced a growing disconnect from the lamp's messages as the study progressed: ``The longer it was learning from us, the more gibberish the poems became'' (P8), ``I think we attribute it to all the poetry being dark'' (P9). 

The decision to include AI technologies brings with it an inherent unpredictability; with increased autonomy and agency the tracing of agencies across these human-machine relationships becomes increasingly blurred and complex. The study's findings illustrate the variety and nuance in experience across participants and households, which speaks to this unpredictable nature, and further motivating the turn to diffractive methodologies.

Across all households, the relationship developed by participants towards the lamp over the duration of the study was inextricably linked to the relationships participants have with each other and with the researchers/designers. Understanding the lamp's \emph{becoming} can only be done through the tracing of these relationships.
At the conclusion of the study, H1 and H2 in particular expressed strong reluctance in surrendering the lamp, with some participants citing the daily poetry as the reason, while others set to miss the lamp's presence in their home. What participants are reluctant to let go of may be not the lamp as a solitary object, but the way in which it enriches the wider network of relationships that surround it.

\section{Discussion and Future Work}

\begin{quote}
    ``This is not to say that if practitioners construct their research from messy diffractive methodologies, they are not accountable for their contributing outcomes. Rather, this underscores the importance and ethics of the journey not over but alongside the destination because we too are reconfigurable through our praxis in ways possibly more consequential than any other.''
\flushright --- \citeauthor{key2022feminist} \cite{key2022feminist}
\end{quote}

\changed{The methodological approach taken in the preceding evaluation was a departure from traditional analyses in that it was not attempting to reach a singular conclusion or set of findings, nor to validate a particular hypothesis (i.e.~was \textit{Message Ritual} successful or not?). The results emerging from such `messy' methods are difficult to reduce to a single overarching narrative, as (naturally) each participant had a unique personal experience with the lamp. But more importantly, answering a hypothesis or question (`Yes, it was successful' or `no, it wasn't successful') does not contribute to the goal of understanding the possibilities of what the artefact could become. We found that by considering the multiplicity of experiences (both their similarities and differences) and attempting to understand the configurations that led to those experiences, the methodology assists in identifying the potential value and risk brought about by the artefact in question.}

\changed{In lieu of a singular conclusion, the diffractive analysis led us, as designers, to better inform our conceptualisation of the artefact. The lamp is foremost a social object, rather than an solitary artefact, a mere technological device or tool.} 
A number of participants described the lamp as a welcome addition to their life, easily fitting in with existing routines and rituals of the household. The moments of structured time carved out by the lamp seemed to offer comfort in the particularly stressful time of pandemic lockdown, along with greater connection in the midst of social isolation. As noted by a number of participants, it is rare in the current state of society that they find quiet moments to reflect. We found that this recollection of memories, thoughts, and feelings not only supports memory, but in remembering we are also engaging with the authoring and re-authoring of our narrative identity. \MR provided participants with the space in which to remember, along with the language and concepts to weave those memories into a broader narrative. The feelings experienced towards the lamp are difficult to fully decouple from the feelings towards other household members and the researchers. Feelings of trepidation, responsibility, guilt, affection, and attachment appear to be distributed across the broader assemblage. This is not to diminish the lamp's agency in such phenomena, but rather to acknowledge that such an artefact cannot be fully decoupled from the material conditions in which it was conceived, and in which it is studied.

\changed{The diffractive analysis furthermore reoriented our approach as researchers to evaluation of AI technologies. Bringing attention to the full spectrum of participant experiences led to a keener awareness of the ethical considerations surrounding such open-ended systems.}
The introduction of AI technologies to deeply intimate and personal spaces carries with it an inherent risk, particularly when those technologies exert their own generative autonomy and agency. We attempted to mitigate these risks through transparency about the capabilities of the system, and informing participants about the open nature of interpretation of machine generated poetry. Ethical accountability involves not only an increased transparency of the the underlying technical processes, but similar requirements extend to the wider network of those who enable \changed{and proliferate} these technologies.

Lastly, the analysis provided valuable insights into many aspects of the current design, including what might be changed or improved in future iterations of \MR. Feedback from participants on extra functionality that they would like to see in the current design was particularly informative. Several participants wished that the lamp could be more ``portable'' (P1, P3, P5) such that it could catch conversations occurring in other rooms of the house. Extending to a network of lamps---potentially of varying sizes and shapes---that can live in different rooms or easily portable would be one way of addressing this need. Other participants asked for the mapping between their language and the lamp's poetry to be more complex (P6, P7), incorporating more of their language (P8, P9), and potentially at different stages of the poetry rather than only at the beginning. We are currently developing more advanced mapping techniques such as the use of n-grams, and dynamically incorporating participant's language in the poetry generation to improve the mapping between conversation and poetry produced. 

This preliminary evaluation informed both future design and evaluation improvements for \MR. The exploratory nature, limited duration \changed{and relatively narrow age demographic} of the current study invites a more detailed, longitudinal study to further understand the variety of ways the lamp can become with others. This would include a more diverse range of households, relationships, and demographics---where participants live with the lamp for extended periods (months, or even years)---to better understand if and how the lamp might become an ongoing ritual. Other factors, such as homes with multi-lingual conversations and transient household members will also be considered (for technical and logistic reasons they were not considered for this study). 

\section{Conclusion}
\label{s:conclusion}
\changed{We have shown how the application of entanglement methodologies to the design, development and evaluation of a novel technological artefact can benefit our understanding of the wider implications and considerations that emerging technologies facilitate. We draw from posthumanist theories to inform the conceptual basis, the design considerations, and the methodological approach to understanding how an artefact \emph{becomes with} the world. This design process encourages consideration of the multiple implicated relationships as they evolve over time. It entails designing not for a specific outcome or function, but in anticipation of how different human-machine relationships might materialise. The presented evaluation favours a diffractive approach to understanding qualitative data; not focusing on whether the design was `successful', but on how different configurations lead to the artefact becoming different things \cite{Frauenberger_2020}.}

\MR is an domestic, ambient and intelligent system that encourages the framing and re-framing of memories through generated poetics. It does this by engaging members of a household through daily rituals centred around language, both shared and private, carried out with intention, and imbued with meaning.
Han doesn't necessarily call for a return to ritual, rather simply that ``rituals serve as a background against which our present times may be seen to stand out more clearly'' \cite[p.vi]{Han2020}. In this paper, we have offered an example of how ritual interaction is contrasted with many of the current ways we relate to technology. With this work, we are not attempting to return to, or reproduce the ritual practices observed across societies past and present, but instead to offer new rituals that produce shared meaning in human and non-human assemblages; new rituals in an increasingly posthuman society.

Through diffractive analysis, we outlined the experiences and perceptions of participants living with lamp; tracing the entanglements between the lamp, participants, and researchers. By diffracting interview data with theory, and again with personal reflection, we teased out our understanding of the lamp's \emph{becoming}, \changed{identifying \emph{some} of the possible roles the \MR system can take: Lamp as diary, as confidant, as new housemate, as morning ritual, as conversation starter, as consumer product, as a technological extension of its (human) designers.}

\changed{In the wake of our analysis, we must ask ourselves as researchers and designers, are these the kind of relationships we \emph{want} to foster?
A diffractive approach allowed us to trace a spectrum of participant experiences to better identify the potential value and risk of an open-ended system such as \MR. In doing so, ethical considerations were brought the fore, and led to an increased sense of accountability towards the introduction of AI technologies into domestic contexts.} We hope that the novel approach presented in this paper inspires further interest in the development of entanglement methodologies in HCI research.

\begin{acks}
This research was supported by Australian Research Council Grants FT170100033 and DP220101223.
\end{acks}

\bibliographystyle{ACM-Reference-Format}
\bibliography{main}

\appendix

\end{document}